\documentclass[12pt,preprint]{aastex} 

\IfFileExists{srcltx.sty}{\usepackage[active]{srcltx}} 

\begin{document}

\title{Searching for dark matter in X-rays: how {\em not} to check the
dark matter origin of a spectral feature}

\author{Alexander Kusenko\altaffilmark{1,2}, Michael
Loewenstein\altaffilmark{3,4}}

\altaffiltext{1}{Department of Physics and Astronomy, University of
California, Los Angeles, CA 90095-1547, USA}
\altaffiltext{2}{Institute for the Physics and Mathematics of the Universe,
University of Tokyo, Kashiwa, Chiba 277-8568, Japan}
\altaffiltext{2}{Department of Astronomy, University of Maryland,
College Park, MD.}
\altaffiltext{3}{CRESST and X-ray Astrophysics Laboratory NASA/GSFC,
Greenbelt, MD.}


\begin{abstract}
In a recent preprint entitled {\em Searching for dark matter in
X-rays: how to check the dark matter origin of a spectral feature} [arXiv:1001.0644], 
the authors have claimed that some archival X-ray data could be used to
rule out dark matter in the form of 5-keV sterile neutrinos at the
level of 20$\sigma$.  Unfortunately, the limit was derived incorrectly.  
We point out the shortcomings of this analysis and show that
the tentative detection of a spectral feature consistent with a 5-keV
sterile neutrino is not in contradiction with existing limits. Future
observations of dwarf spheroidal galaxies will test this hypothesis.
\end{abstract}



Sterile neutrinos with masses of several keV can make up all or most
of dark matter~\citep{dw94}, can explain the observed velocities of
pulsars~\citep{ks97,fkmp03,fk06,kmm08}, and can play a role in the
formation of the first stars and in other astrophysical
phenomena~\citep{bk06,K09}. The first dedicated search for relic
sterile neutrinos using {\em Suzaku}~\citep{lkb09} and {\em
Chandra}~\citep{lk09} X-ray telescopes has produced some reliable new
limits, as well as a tentative detection of a spectral feature
consistent with a decay line of a 5~keV sterile neutrino.

However, \cite{briwrsh10} have claimed that this detection can be
excluded at the level of 20$\sigma$ based on the archival data from
observations of M31. The X-ray data used by \cite{briwrsh10} are the
same data that \cite{wat06} have used to produce an exclusion limit
(which is not in disagreement with our result). Obviously, it would be
quite remarkable if these archival data turned out to be better suited
to search for sterile neutrinos than the ongoing and planned dedicated
X-ray observations~\citep{lk09,lkb09}. We examine this claim, and we find
it to be unfounded.

The strength of the expected signal depends on the amount of dark
matter in the field of view. The strongest results of \cite{briwrsh10}
were obtained based on the central region of M31.  More specifically,
they considered a ring with inner and outer radii of 5\arcmin and
13\arcmin~that correspond to 1.1~kpc and 3~kpc, respectively, at the
distance of M31. To obtain the amount of dark matter in this part of
their field of view, \cite{briwrsh10} considered several smooth
profiles hypothesized in the literature, each of which gives a
reasonable, but by no means unique, fit to the data. Of these profiles
they chose the one that produces the lowest projected mass in this
annulus, $1.1\times 10^{10}~M_\odot$, corresponding to a column mass
density of 606~$M_\odot {\rm pc}^{-2}$.  Their alleged exclusion limit
is based on the assumption that at least this much of dark matter
falls within the selected region of the field of view. If the amount
of dark matter is lower by a factor of two or more, their claim is
weakened or invalidated, as the expected signal-to-noise ratio
decreases accordingly.

The problem with the analysis of \cite{briwrsh10} is that the smooth
profiles they chose to consider do not represent the full range of
uncertainty in the dark matter content in their field of view. The
mass density in the central region of M31 is known to be dominated by
baryonic matter. As emphasized by \cite{klypin02}, a dark matter
profile inferred from cold dark matter simulations is not applicable
to the actual dark matter distribution in the central 3~kpc of M31,
because the interactions of dark matter with baryons are expected to
facilitate the angular momentum transfer and expulsion of dark matter
from the central region. The presence of a rotating bar in M31 implies
that the central density of dark matter should be sufficiently small to
avoid causing the dynamical friction that would interfere with the
rotating bar \citep{weinberg85,debatt00}. The spherical profiles may
be incorrect for fitting the data because the dark matter distribution
in the central region can be triaxial~\citep{binney00}. Finally,
sterile neutrinos have different clustering properties on small scales
from those of cold dark
matter~\citep{petraki,boyan08a,boyan08b,boyarsky09}, and there are no
reliable calculations of dark matter profiles that take into account
both the free-streaming of warm dark matter and the effects of the
baryons.

Even aside from all of these caveats, the simplest spherical profiles
giving the best fit to observational data require much less dark
matter in the relevant region than what was claimed by \cite{briwrsh10}, who fail to
consider some of the most recent work on the mass distribution in M31.
The baryonic mass, which dominates in the center of M31, 
itself is composed of (at least) two (bulge and disk) components with
distinct mass-to-light (M/L) ratios that are not known {\it a priori},
but must be estimated from optical spectra under various simplifying
assumptions about their respective stellar populations. \cite{ccf}
present models that include stellar mass-to-light ratios based on
stellar population synthesis models that have average column mass
densities in the 1.1-3 kpc annulus $\sim 1.5$ times lower than the
``minimum'' value in \cite{briwrsh10} (while noting the lack of
axisymmetry in the inner regions and presence of a central velocity
dip in the HI rotation curve, and stating that that all their models
``fail to reproduce the the exact shape of the rotation curve''). A
new analysis of M31 based on the recent deep, full-disk 21-cm imaging
survey~\citep{cor09} shows that the best-fit profile is achieved using
the \cite{burk95} parametrization with the scale parameter
$R_B=77$~kpc (for which $\chi^2=0.81$ indicates a good fit to the
data). For this profile, we obtain $3.4\times 10^{9}~M_\odot$ for the
total projected dark matter mass in the 5\arcmin-13\arcmin~region of
M31. When~\cite{cor09} impose a constraint on the virial mass of M31,
their best fit corresponds to $R_B=28$~kpc, in which case we obtain
$2.9\times 10^{9}~M_\odot$ for the total projected dark matter mass in
this region, corresponding to a column density of 128~$M_\odot {\rm
pc}^{-2}$. In reality, the dark matter content of the region in
question could be even smaller than the value obtained from the
best-fit profile. In these models, the stellar mass-to-light ratios
are higher than in \cite{ccf}; and indeed, a recent study
\citep{saglia10} supports such a higher M/L in the bulge that dominates
the stellar mass in the central regions. The M31 rotation curve is
clearly consistent with a dark matter mass and column density that are
factor $>4$ below what \cite{briwrsh10} have claimed to be the minimal
mass and the minimal density.

Since the minimal dark matter mass in the region chosen by
\cite{briwrsh10} is at least 4 times smaller than the value they have
assumed, the expected minimal intensity of the  decay line should be
scaled down by at least a factor of 4. Such a properly scaled feature
 falls well below the Galactic and Cosmic X-ray backgrounds, and
it is completely swamped by the noise, making impossible the derivation
of any robust exclusion limit. At these levels, unresolved hard X-ray
emission from low-mass X-ray binaries and other stellar (and possibly
interstellar) sources come into play \citep{bg08}, which may complicate
the expected shape of the $>2$ keV continuum.  These additional
components are completely absent in the ultra-faint dwarf spheroidals
\citep{lkb09}, such as Willman~1. Unlike the case of M31, dark matter
predominates in these systems, and the dark matter distribution is
fully determined by kinematics without the uncertainties introduced by
the presence of multiple mass components.

For the reasons stated above, M31 is not as good a target for future
observations as dwarf spheroidal galaxies. The ongoing and planned
observations of dwarf spheroidal galaxies give the best opportunity to
confirm or rule out 5~keV sterile neutrinos as a dark matter
candidate.

Support for this work was provided by the National Aeronautics and
Space Administration through {\it Chandra} Award Numbers G08-9091X and
G09-0090X issued by the Chandra X-ray Observatory Center, which is
operated by the Smithsonian Astrophysical Observatory for and on
behalf of the National Aeronautics Space Administration under contract
NAS8-03060. The work of AK was supported in part by DOE grant
DE-FG03-91ER40662 and NASA ATFP grant NNX08AL48G.


{}

\clearpage

\end{document}